\documentstyle[12pt]{article}

\title{The Quantum Theory of General Relativity \\ at Low Energies}

\author{John Donoghue\\ [5mm]
 Department of Physics and Astronomy\\ University of
Massachusetts Amherst, MA 01003 U.S.A.}
\date{ }
\begin{document}
\begin{titlepage}
\maketitle
\begin{abstract}
In quantum field theory there is now a well developed 
technique, effective field theory, which allows one to obtain 
low energy quantum predictions in ``non-renormalizable'' theories, 
using only the degrees of freedom and interactions appropriate 
for those energies. Whether or not general relativity is truly 
fundamental, at low energies it is automatically described 
as a quantum effective field theory and this allows a consistent 
framework for quantum gravity at ordinary energies. I briefly 
describe the nature and limits of the technique.
\end{abstract}
{\vfill gr-qc/9607039  Talk presented at Journees Relativistes 96,
Ascona, Switzerland, May 1996, to be published in Helvetia Physica Acta.}
\end{titlepage}

Effective field theory is a calculational technique in quantum field
theory which has become fully developed within the past decade[1]. It is
now in everyday use in a variety of contexts. Anyone who cares about
quantum field theory should be familiar with the methods and insights
of effective field theory.
   
The goal of this talk is to convince you that a consistent quantum
theory of general relativity exists at energies well below the Planck
mass, and that it is necessarily of the form that we call effective 
field theory. Given all the work that has gone into quantum gravity, 
I feel that this is a significant result. Indeed, the gravitational 
effective field theory[2,3] is likely the full quantum content of 
pure general relativity.
 
The gravitational effective field theory is a completion of the
program started by Feynman and DeWitt[4], 'tHooft and Veltman[5] and many
others. The previous work focused on the divergence structure and the
problems at high energy. What the effective field theory techniques 
do is to shift
the focus to low energy, which is the reliable part of the theory. 
The low energy particles and interactions lead to quantum effects
which are distinct from whatever physics is going on at very high
energies. For
example, the long range quantum correction to the gravitational
potential is determined by the low energy interactions of the massless
particles in the theory (gravitons, photons, and neutrinos) and is
reliably calculable. For a
particular definition of the potential, it has the form[3,6]
\begin{equation}
V_{1pr}(r) = -{Gm_1m_2 \over r}\left[ 1 - {135 + 2 N_\nu \over
30 \pi^2 }{G \hbar \over r^2 c^3} +\ldots \right]
\end{equation}
\noindent where $N_\nu$ is the number of massless neutrinos. I will
here focus on the general nature and limits of the gravitational
effective field theory.

First let's describe effective field theory in general. Once you
understand the basic ideas it is easy to see how it applies to
gravity. The phrase ``effective'' carries the connotation of a low
energy approximation of a more complete high energy theory. However,
the techniques to be described don't rely at all on the high energy
theory. (Moreover, even if you believe that general relativity is
exact and fundamental at all scales, these techniques are still
appropriate at low energy.) It is perhaps better to focus on a second
meaning of ``effective'', ``effective'' $\sim$ ``useful'', which
implies that it is
the most effective thing to do. This is because the particles and
interactions of the effective theory are the useful ones at that
energy. An ``effective Lagrangian'' is a local Lagrangian which
describes the low energy interactions. There is an old fallacy that
effective Lagrangians can be used only at tree level. This sometimes
still surfaces despite general knowledge to the contrary. ``Effective
field theory'' is more than just the use of effective Lagangians. It
implies a specific full field-theoretic treatment, with loops,
renormalization etc. The goal is to extract the full quantum effects
of the particles and interactions at low energies.

The key to the separation of high energy from low is the uncertainty
principle. When one is working with external particles at low energy,
the effects of virtual heavy particles or high energy intermediate
states involve short distances, and hence can be represented by a
series of local Lagrangians. This is true even for the high energy
portion of loop diagrams. In contrast, effects that are non-local,
where the particles propagate long distances, can only come from the
low energy part of the theory. From this distinction, we know that we
can represent the effects of the high energy theory by the most
general local effective Lagrangian. The second key is the energy
expansion, which orders the infinite number of terms within this most
general Lagrangian in powers of the low energy scale divided
by the high energy scale. To any given order in this small parameter,
one needs to deal with only a finite number of terms (with
coefficients which in general need to be determined from experiment).
The lowest order Lagrangian can be used to determine the propagators
and low energy vertices, and the rest can be treated as perturbations.
When  this theory is quantized and used to calculate loops, the usual
ultraviolet divergences will share the form of the most general
Lagrangian (since they are high energy and hence local) and can be
absorbed into the definition of renormalized couplings. There are
however leftover effects in the amplitudes from long distance
propagation which are distinct from the local Lagrangian and which are
the quantum predictions of the low energy theory.

This technique can be used in both renormalizable and
non-renormalizable theories, as there is no need to restrict the
dimensionality of terms in the Lagrangian. (Note that the terminology
is bad: we {\it are} able to renormalize non-renormalizable theories!) 
One example, which is amusing to describe to die-hard 
loyalists who insist on the
renormalizable paradigm, is Heavy Quark Effective Theory[7]. Here we have
a perfectly good renormalizable field theory (QCD), yet we choose to
turn it into a non-renormalizable effective field theory by a field
redefinition which isolates the most important heavy quark degree of
freedom. This is the effective thing to do because the properties of
heavy quarks become readily apparent and the difficult parts of QCD
are contained in a few universal parameters or functions. The
effective field theory which is most similar to general relativity is
chiral perturbation theory[8], which describes the theory of pions and
photons which is the low energy limit of QCD. The theory is highly
nonlinear, with a lowest order Lagrangian which can be written with
the exponential of the pion fields
\begin{eqnarray}
{\cal L} & = & {F_\pi^2 \over 4} Tr\left( \nabla_\mu U \nabla^\mu
U^\dagger \right) + {F_\pi^2 m_\pi^2 \over 4} Tr\left( U+U^\dagger
\right) \nonumber \\
U & = & exp \left(i {\tau^i \pi^i(x) \over F_\pi} \right) \ \ ,
\end{eqnarray}
\noindent with $\tau^i$ being the SU(2) Pauli matrices and $F_\pi
=92.3 MeV$ being a dimensionful coupling constant. This theory has
been extensively studied theoretically, to one and two loops, and
experimentally. There are processes which clearly reveal the
presence of loop diagrams. In a way, chiral perturbation theory is the
model for 
a complete non-renormalizable effective field theory in the same way
that QED serves as a model for renormalizable field theories.

At low energies, general relativity automatically behaves in the way
that we treat effective field theories. This is not a philosophical
statement implying that there must be a deeper high energy theory of which
general relativity is the low energy approximation (however, more on
this later). Rather, it is a practical statement. Whether or not 
general relativity is truly fundamental, the low energy quantum
interactions must behave in a particular way because of the nature of
the gravitational couplings, and this way is that of effective field
theory. 

The Einstein action, the scalar curvature, involves two
derivatives on the metric field. Higher powers of the curvature,
allowed by general covariance, involve more derivatives and hence the
energy expansion has the form of a derivative expansion. (The
renormalized cosmological constant is small on ordinary scales and so
I neglect it, although it could possibly be treated as a perturbation
itself, as the pion mass is treated in chiral perturbation theory.)
The higher powers of the curvature in the most general Lagrangian do
not cause problems when treated as low energy perturbations.[9]
The Einstein action is in fact readily quantized, using gauge-fixing
and ghost fields ala Feynman, DeWitt, Faddeev, Popov[4]. The background
field method used by 'tHooft and Veltman[5] is most beautiful in this
context because it allows one to retain the symmetries of general
relativity in the background field, while still gauge-fixing the
quantum fluctuations. The dimensionful nature of the gravitational
coupling implies that loop diagrams (both the finite and infinite
parts) will generate effects at higher orders in the energy
expansion[10]. The one and two loop counterterms for graviton
loops are known[5,11] and go into 
the renormalization of the coefficients in
the Lagrangian. However, these are not really predictions of the
effective theory. The real action comes at low energy.

How in practice does one separate high energy from low? Fortunately,
the calculation takes care of this automatically, although it is
important to know what is happening. Again, the main point is that the
high energy effects share the structure of the local Lagrangian, while
low energy effects are different. When one completes a calculation,
high energy effects will appear in the answer in the same way that the
coefficients from the local Lagrangian will. One cannot
distinguish these effects from the unknown coefficients. However, low
energy effects are anything that has a different structure. Most
often the distinction is that of analytic versus non-analytic in
momentum space. Analytic expressions can be Taylor expanded in the
momentum and therefore have the behavior of an energy expansion, much
like the effects of a local Lagrangian ordered in a derivative
expansion. However, non-analytic terms can never be confused with the
local Lagrangian, and are intrinsically non-local. Typical
non-analytic forms are $\sqrt{-q^2}$ and $\ln(-q^2)$. These are always
consequences of low energy propagation. 

A conceptually simple (although calculationally difficult) example is
graviton - graviton scattering. This has recently been calculated to
one-loop in an impressive paper by Dunbar and Norridge[12] using string
based methods. Because the reaction involves only the pure gravity
sector, and $R_{\mu\nu} = 0$ is the lowest order equation of motion,
the result is independent of any of the four-derivative terms that can
occur in the Lagrangian ($R^2$ or $R_{\mu\nu}R^{\mu\nu}$)[5]. Thus the
result is independent of any unknown coefficient to one loop order.
Their result for the scattering of positive helicity gravitons is
\begin{eqnarray}
{\cal{A}}(++\to ++) & = & 8\pi G {s^4 \over stu} \left\{ \right. 1
\nonumber  \\
 & + & {G \over \pi} \left[  \right. ~\left( t \ln ({-u\over \delta}) 
\ln ({-s
\over \delta}) + u \ln ({-t\over \delta }) \ln ({-s\over \delta}) 
+ s \ln ({-t \over \delta}) \ln ({-u\over \delta}) \right)
\nonumber \\
 & + & \ln ({t \over u}){tu(t-u)\over 60 s^6}\left( 341(t^4+u^4) +1609
(t^3u + u^3t) +2566 t^2u^2 \right)  \nonumber \\
 & + & \left( \ln^2 ({t\over u}) + \pi^2\right){tu(t+2u)(u+2t)\over
2s^7} \left(2t^4 + 2u^4 +2t^3u +2u^3t - t^2u^2\right)   \nonumber \\
 & + & {tu \over 360 s^5} \left( 1922(t^4 + u^4) +9143(t^3u+u^3t) 
+14622 t^2u^2 \right) \left. ~ \right]
\left. ~ \right\} \ \ 
\end{eqnarray}
\noindent where $s=(p_1+p_2)^2$, $t=(p_1-p_3)^2$, $u=(p_1-p_4)^2$,  
($s+t+u=0$)
and where I have used $\delta$ as an infrared cutoff. One sees the
non-analytic terms in the logarithms. Also one sees the nature of the
energy expansion in the graviton sector - it is an expansion in $G
E^2$ where E is a typical energy in the problem. I consider this
result to be very beautiful. It is a low energy theorem of quantum
gravity. The graviton scattering amplitude {\it must} behave in this
specific fashion no matter what the ultimate high energy 
theory is and no matter
what the massive particles of the theory are. This is a rigorous
prediction of quantum gravity.

The other complete example of this style of calculation is the long
distance quantum
correction to the gravitational interaction of two masses. Again the
result is independent of any unknown coefficients in the general
matter and gravity Lagrangian, because the effect of such analytic
terms is to lead to a short range 
delta-function interaction[3,13]. Only the
propagation of massless fields can generate 
the nonanalytic behavior that
yields power-law corrections in 
coordinate space[2,3]. Since the low energy
couplings of massless particles are determined by Einstein's theory,
these effects are also rigorously calculable. Besides classical
corrections[14], one obtains the true quantum correction as quoted in the
introduction above. Note that this calculation is the first to provide
a  quantitative answer to 
the question as to whether the effective gravitational coupling
increases or decreases at short distance due to quantum effects. While
there is some arbitrariness in what one defines to be $G_{eff}$, it
must be a universal property (this eliminates from consideration
the Post-Newtonian classical correction which depends on the external
masses) and must represent a general property of the theory. The
diagrams involved in the above potential are the same ones that go
into the definition of the running coupling in QED and QCD and the
quantum corrections are independent of the external masses. If one
uses this gravitational interaction to define a running coupling one
finds
\begin{equation}
G_{eff}(r) = G \left[ 1 - {135 + 2 N_\nu \over
30 \pi^2 }{G \hbar \over r^2 c^3}  \right]
\end{equation}
\noindent The quantum corrections {\it decrease} the strength of
gravity at short distance, in agreement with handwaving expectations.
(In pure gravity without photons or massless neutrinos, 
the factor $135 +2N_\nu$ is replaced by $127$.)
An alternate definition including the diagrams calculated in [6] has
a slightly different number, but the same qualitative conclusion. The
power-law running, instead of the usual logarithm, is a consequence of
the dimensionful gravitational coupling. 
 
These two results do not exhaust the predictions of the effective
field theory of gravity. In principle, any low energy gravitational
process can be calculated[15]. The two examples above have been
particularly nice in that they did not depend on any unknown
coefficients from the general Lagrangian. However it is not a failure
of the approach if one of these coefficients appears in a particular
set of amplitudes. One simply treats it as a coupling constant,
measuring it in one process (in principle) and using the result in the
remaining amplitudes. The leftover structures aside from this
coefficient are the low energy quantum predictions.

The effective field theory techniques can be applied at low energies
and small curvatures.
The techniques fail when the energy/curvature reaches the Planck
scale. There is no known method to extend such a theory to higher
energies. Indeed, even if such a technique were found, the result
would likely be wrong. In all known effective theories, new degrees of
freedom and new interactions come into play at high energies, and to
simply try to extend the low energy theory to all scales is the wrong
thing to do. One needs a new enlarged theory at high energy. However,
many attempts to quantize general relativity ignore this distinction
and appear misguided from our experience with other effective field
theories. While admittedly we cannot be completely sure of the high
energy fate of gravity, the structure of the theory itself hints very
strongly that new interactions are needed for a healthy high energy
theory. It is likely that, if one is concerned with only pure general
relativity, the effective field theory is the full quantum content of
the theory. 

It is common to hear that gravity is different from all our other
theories because gravity and quantum mechanics do not go together,
that there is no quantum theory of gravity. This is not really the
case, as there is no conflict between gravity and quantum mechanics
at low energy. We also expect that all of our other theories, despite
being renormalizable, are modified by new interactions at high energy.
Nevertheless, we are content to make predictions with them in the 
region where they are valid. While gravity at low energies has a
somewhat different structure than other theories, it is not that a
quantum theory does not exist. 
Rather the more accurate statement is that the quantum theory of
gravity reveals itself as an effective field theory at low energies
and signals that we need a more elaborate theory at high energies.

\end{document}